# Ermittlung von Verwundbarkeiten mit elektronischen Ködern

Maximillian Dornseif

Institut für Strafrecht, Universität Bonn

Felix C. Gärtner  Thorsten Holz

Lehr- und Forschungsgebiet Informatik 4, RWTH Aachen

**Abstract:** Als elektronische Köder (*honeypots*) bezeichnet man Netzwerkressourcen, deren Wert darin besteht, angegriffen und kompromittiert zu werden. Oft sind dies Computer, die keine spezielle Aufgabe im Netzwerk haben, aber ansonsten nicht von regulären Rechnern zu unterscheiden sind. Köder können zu Köder-Netzwerken (*honeynets*) zusammengeschlossen werden. Sie sind mit spezieller Software ausgestattet, die die Forensik einer eingetretenen Schutzzielverletzung erleichtert. Durch die Vielfalt an mitgeschnittenen Daten kann man deutlich mehr über das Verhalten von Angreifern in Netzwerken lernen als mit herkömmlichen forensischen Methoden. Dieser Beitrag stellt die Philosophie der Köder-Netzwerke vor und beschreibt die ersten Erfahrungen, die mit einem solchen Netzwerk an der RWTH Aachen gemacht wurden.

## 1 Einleitung

„Angenommen", sagte er [Winnie Pu] zu Ferkel, „*du* willst *mich* fangen, wie würdest Du das machen?"

„Tja", sagte Ferkel, „ich würde es so machen: Ich würde eine Falle bauen, und ich würde einen Topf Honig in die Falle stellen, und Du würdest den Honig riechen, und Du würdest in die Falle gehen, und ..."

[Mi87, S. 64]

Der englische Begriff *honeypot* bezeichnet für gewöhnlich einen Gegenstand, von dem eine gewisse Attraktivität ausgeht, die bestimmte, nicht nur tierische, Interessenten anzulocken vermag [Lo01]. Sie eignen sich demnach als Köder, um Aufmerksamkeit auf einen Gegenstand zu lenken. Wissenschaftler haben neuerdings begonnen, das Prinzip der Köder auch im Bereich der IT-Sicherheit anzuwenden. Hier werden *elektronische Köder* ausgelegt, um das Verhalten von Angreifern leichter zu studieren. Elektronische Köder sind Netzwerkressourcen (Computer, Router, Switches), deren Wert darin besteht, angegriffen und kompromittiert zu werden [Sp02]. Die Köder haben keine spezielle Aufgabe im Netzwerk, sind aber ansonsten nicht von regulären Komponenten zu unterscheiden. Sie sind mit spezieller Software ausgestattet, die die anschließende Forensik eines Angriffs deutlich erleichtert. Im Gegensatz zu einer herkömmlichen forensischen Untersuchung erlauben beispielsweise gezielte Veränderungen im Betriebssystem das direkte Mitschneiden aller Aktivitäten eines Angreifers. Durch die Vielfalt der so gewonnenen Daten kann man schneller und genauer dessen Angriffswege, Motive und Methoden erforschen. Auch kann man die benutzten Angriffswerkzeuge, die normalerweise nach erfolgter Kompromittierung gelöscht werden, sofort sicher stellen. Schon relativ früh hatte es Versuche gegeben, mit entsprechenden Modifikationen die Aktivitäten von Angreifern aufzuzeichnen [St88, Ch90]. Diese Versuche erfolgten jedoch nicht systematisch, sondern aus der aktuellen Situation heraus, in der ein Einbruch beobachtet wurde.

Der *honeypot*-Ansatz erscheint sowohl aus praktischer als auch aus theoretischer Sicht interessant für den Bereich der IT-Sicherheit. Aus praktischer Sicht erlauben detaillierte Informationen über das Verhalten von böswilligen „Hackern" (den sogenannten *blackhats*), die Abwehrmaßnahmen in unterschiedlichen Umgebungen effizienter und effektiver zu gestalten. Effizienter werden Abwehrmaßnahmen dadurch, dass man sich auf *relevante* Angriffe konzentrieren kann (Netzwerke von *honeypots*, so genannte *honeynets*, können hierzu empirische Beiträge liefern). Effektiver werden Abwehrmaßnahmen, indem mit Hilfe der *honeypots neue* Schwachstellen und Verwundbarkeiten entdeckt und schnell analysiert werden können.

Aus theoretischer Sicht bilden *honeypots* eine interessante Instanz des *dual use*-Prinzips der IT-Sicherheit: Man bekämpft Angreifer mit ihren eigenen Methoden. Die Fähigkeiten und die Werkzeuge, die ein Verteidiger anwendet, unterscheiden sich kaum von denen der Angreifer. So entstand beispielsweise die *honeynet*-Software Sebek [Th03b] aus einem verbreiteten Hackerwerkzeug (einem *rootkit*). Außerdem versucht der Betreiber eines *honeypot* genau das, was ein erfahrener Angreifer ebenfalls tun würde, nämlich: seine Existenz im System mit allen Mitteln zu verbergen.

Am Lehr- und Forschungsgebiet Informatik 4 (Verlässliche Verteilte Systeme) der RWTH Aachen wird im Rahmen eines durch die DFG geförderten Projektes der Einsatz von *honeypots* zur Gefahrenabwehr in der IT-Sicherheit untersucht. Dieser Beitrag berichtet von den ersten Erfahrungen, die beim Aufsetzen eines elektronischen Köder-Netzwerks gesammelt werden konnten. Die hierbei gewonnenen Einsichten deuten darauf hin, dass *honeynets* eine wertvolle Unterstützung für die IT-Sicherheitsforschung sein können.

Abschnitt 2 erläutert Ursprünge, Ziele und Erfolge bisheriger *honeynet*-Forschungen. In Abschnitt 3 wird der Aufbau des *honeynet* an der RWTH Aachen näher beschrieben. Daran schließt sich in Abschnitt 4 ein erster Bericht über die bisher gemachten Erfahrungen an. Die ethischen und rechtlichen Aspekte beim Betrieb eines *honeynet*, die in Deutschland beachtet werden müssen, werden in Abschnitt 5 näher beschrieben. Abschnitt 6 gibt einen Überblick über die geplanten Arbeiten in diesem Themengebiet.

## 2 Das globale Honeynet-Projekt

### 2.1 Motivation

Unter dem Namen *The Honeynet Project* [hon04a] haben sich Forscher aus der ganzen Welt zusammengeschlossen, die sich mit dem Thema IT-Sicherheit beschäftigen. Ziel des Projekts ist, die Programme, Vorgehensweisen und Motive von böswilligen Angreifern (*blackhats*) kennenzulernen und das so gewonnene Wissen allen Interessierten zur Verfügung zu stellen. Das Augenmerk liegt vor allem auf folgenden Punkten:

- **Bewusstsein schaffen**

  Es soll bewusst gemacht werden, in welchem Ausmaß Bedrohungen und Sicherheitslücken in Netzwerken bestehen. Dazu werden Computersysteme „echten" Angriffen aus dem Internet ausgesetzt und Daten über die Geschwindigkeit und den Grad der Schwierigkeit (oder vielmehr der Leichtigkeit) einer Kompromittierung veröffentlicht.

- **Lernen über bekannte Angriffswege**

  Durch die Erfahrungen mit *honeynets* kann man genaue Rückschlüsse ziehen auf die Art und Weise, wie Angreifer vorgehen, sowie auf die Werkzeuge, die sie dabei verwenden. Diese Informationen erlauben einem Systemadministrator, sich effizient gegen Verletzlichkeiten abzusichern. Effizient bedeutet hierbei, dass man den Einsatz von Ressourcen für die Gefahrenabwehr gezielter planen kann.

- **Lernen über unbekannte Angriffswege**

  *Honeynets* ermöglichen erstmals, Angreifern bei ihrer Arbeit „über die Schulter zu schauen" und so Bedrohungen und Angriffe in einer neuen Qualität zu erforschen. Sehr leicht läßt sich beispielsweise die zeitliche Reihenfolge rekonstruieren, in der ein Angreifer vorgegangen ist. Dies ist bei herkömmlichen *a posteriori* forensischen Untersuchungen (wenn überhaupt) nur mit großem Aufwand möglich. Außerdem gewähren *honeynets* neue Einblicke in die Motive und Kommunikationswege von Angreifern. In diesem Kontext sollen Methoden und Techniken entwickelt werden, die das Sammeln und das Verbergen von Informationen im Rechnernetz ermöglichen.

- **Aktive Verteidigung**

  *Honeynets* dienen direkt dem Schutz von Netzwerken. Als *soft target* in einem besonders gefährdeten Netz sollen sie wie die Magnesium-Anode beim kathodischen Korrosionsschutz die Angreifer von den schützenswerten Zielen ablenken. Weiterhin wird versucht, Angriffe zu erkennen und diese, für den Angreifer transparent, vom Produktionssystem auf einen *honeypot* umzulenken.

## 2.2 Vorgehensweise

Im Rahmen des Honeynet-Projekts werden Werkzeuge für den Einsatz von *honeypots* entwickelt. Hierzu gehört die spezielle Überwachungssoftware Sebek [Th03b], die ihre Existenz auf dem *honeypot* verbirgt. Sebek ist ein Client/Server-System: Auf dem *honeypot* selbst befindet sich der Sebek-Client, der, für einen Angreifer nicht wahrnehmbar, alle Aktivitäten mitschneidet. Diese Informationen werden (ebenfalls unmerklich) über das Netzwerk an den Sebek-Server geschickt, der die Daten für die Auswertung speichert und aufbereitet.

Der *honeypot* selbst ist in der Regel ein Computersystem, das keine konventionelle Aufgabe im Netzwerk hat und mit dem deshalb in der Regel niemand interagiert. Diese Annahme vereinfacht die Entdeckung eines Angriffes deutlich: Findet eine Interaktion mit diesem System statt und werden dort Pakete empfangen, deutet dies auf ein Sammeln von Informationen, einen Angriff oder einen Kompromittierung des Systems hin. Deshalb wird sämtlicher Netzwerkverkehr zwischen dem *honeypot* und dem übrigen Netz für die spätere Analyse gesammelt.

Ein *honeynet* ist ein Netzwerk von mehreren *honeypots* und stellt also die Verallgemeinerung des Prinzips dar. Hierbei werden meist verschiedene Typen von *honeypots* in einem Netz aufgesetzt, wodurch man Informationen über Angriffe auf verschiedene Plattformen erhalten kann.

## 2.3 Bisherige Erfahrungen

Der Einsatz von *honeynets* hat bereits wertvolle Informationen über die Vorgehensweise und das Verhalten von *blackhats* gebracht. Die verwendeten Programme und Toolkits (insbesondere *rootkits* und verwendete *exploits*) konnten aufgezeichnet und analysiert werden. Beispielsweise wurde im Januar 2002 mittels der aufgezeichneten Daten eines *honeypot* erstmals ein *exploit* gefunden, der eine vorher schon bekannte Schwachstelle in der `dtspcd`-Software (CDE Subprocess Control Service) zum Angriff nutzte [CE02].

Durch das Mitschneiden aller Aktivitäten kann man das typische Vorgehen nach einem erfolgreichen Angriff besser nachvollziehen. Darüber hinaus liefern mitgeschnittene Konversationen im *Internet Relay Chat* (IRC) Informationen über das soziale Verhalten und die Kommunikationswege von Angreifern. Das Azusa Pacific University Honeynet Research Project entdeckte vor kurzem mittels eines *honeynet* ein sogenanntes *botnet*, i.e. ein Netzwerk von kompromittierten Rechnern, auf denen ein IRC-Client installiert ist, über den der jeweilige Computer eingeschränkt ferngesteuert werden kann. Dieses *botnet* hatte eine Größe von mehr als 15 000 Rechnern [Mc03b]. Große *botnets* wie dieses werden häufig eingesetzt, um *Denial of Service*-Angriffe auf IRC-Server oder Webserver durchzuführen und stellen deshalb eine große Bedrohung dar.

Die Untersuchung dieses *botnet* hat zu weiteren interessanten Erkenntnissen über das soziale Verhalten von Angreifern geführt. Die Forscher entdeckten etwa eine Reihe von IRC-Foren (*channels*), in denen organisierter Handel mit Informationen über Kreditkarten stattfand [Mc03a]. Sie fanden automatisierte Tools, mit deren Hilfe Kreditkartenbetrug vereinfacht wurde und erhielten einen Einblick in die Szene.

Außerdem können aus den aufgezeichneten Informationen von *honeynets* Abschätzungen über die Verwundbarkeit von Computersystemen (*Vulnerability Assessment*) und deren erwartete Zuverlässigkeit gewonnen werden [hon04b].

## 3 Architektur an der RWTH

Am Lehr- und Forschungsgebiet Informatik 4 (Verlässliche Verteilte Systeme) an der RWTH Aachen wurde im Januar 2003 ein *honeynet* aufgesetzt, dessen grundsätzlicher Aufbau in Abbildung 1 zu sehen ist.

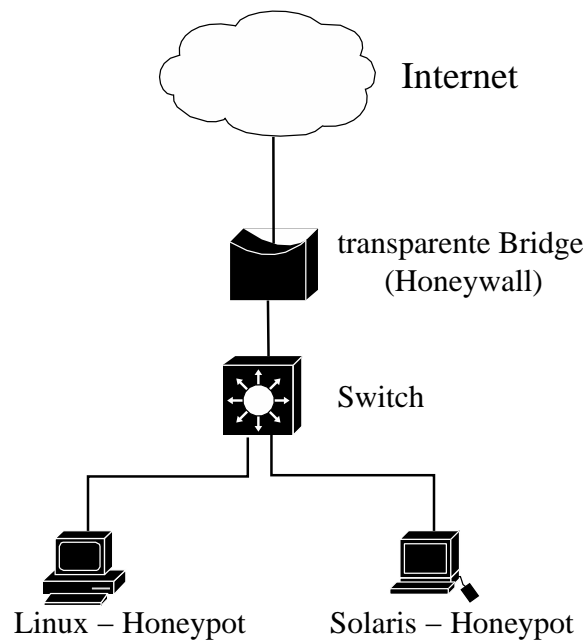

Abbildung 1: Aktueller Aufbau des *honeynet* an der RWTH Aachen.

Auf dem Linux-*honeypot* läuft eine SuSE Linux 8.0 Professional-Installation, die die Dienste HTTP (Apache 1.3.23 inklusive PHP 4.1.0), FTP (vsFTPd 1.0.1) und SSH (OpenSSH 3.0.2p1) anbietet. Außerdem wurde PHP-Nuke [php04] in Version 5.0 sowie MySQL 3.23.53 installiert, um beobachten zu können, ob Angreifer auch spezielle Webapplikationen angreifen. Gegenüber einer Standardinstallation wurden nur wenige Änderungen durchgeführt: Die wichtigste Änderung war die Installation des Sebek-Client. Zudem wurden einige sogenannte *honeytokens* [Th03a] auf dem System hinterlegt. Dies sind verschiedene Arten von Daten (beispielsweise Mails, Tabellenkalkulations- oder verschlüsselte

Dateien), die das Interesse des Angreifers wecken und die den Ablauf der Informationsgewinnung nach einem erfolgreichen Angriff nachzuvollziehbar machen sollen. Desweiteren wurde das System so eingerichtet, dass es einem „normalen" System mit drei Benutzern gleicht.

Der Solaris-*honeypot* ist eine Sun Ultra 1 mit Solaris 8 als Betriebssystem. Dieses System bietet ebenfalls die Dienste HTTP (Apache 1.3.12), FTP (wuftpd-2.6.1) und SSH (OpenSSH 3.02p1) an. Der Sebek-Client sorgt auch hier dafür, dass sämtliche Aktivitäten eines Angreifers auf diesem System aufgezeichnet werden können. Außer der Deaktivierung vieler Dienste des `inetd` (etwa `telnet`, `echo` oder `talk`) wurden keine anderen Modifikationen durchgeführt.

Die beiden *honeypots* sind an einen Switch angeschlossen und von dort aus über eine transparente Bridge mit dem Internet verbunden. Diese transparente Bridge, genannt *honeywall*, ist eine der wichtigsten Komponenten des *honeynet*. Die *honeywall* basiert auf Debian GNU/Linux 3.0r2 und stellt Möglichkeiten zur Steuerung des Datenflusses und Speicherung der aufgezeichneten Informationen bereit. Mittels der `netfilter/iptables`-Funktionalität des Linux-Kernels wird der ausgehende Netzwerkverkehr des *honeynet* limitiert. Beispielsweise dürfen pro Tag nur 15 TCP-Verbindungen je *honeypot* initiiert werden.

Zusätzlich läuft auf der *honeywall* die Snort_inline-Software, ein so genanntes *Intrusion Prevention System*. Dieses sorgt in Verbindung mit `iptables` dafür, dass ein Angreifer nach erfolgter Kompromittierung eines *honeypot* nicht von dort aus noch weitere Systeme angreifen kann. Snort_inline basiert auf Snort, einem quelloffenen *Intrusion Detection System*, und erweitert dessen Funktionalität um Methoden zur Datensteuerung speziell für *honeynets*. Dazu werden die Snort-Regeln so umgeschrieben, dass bekannte Angriffe auf andere Systeme nicht geblockt, sondern modifiziert und somit effektiv verhindert werden. Abbildung 2 zeigt ein Beispiel für eine Snort_inline-Regel, mit der Pakete, die Shellcode für die x86-Architektur enthalten, so verändert werden, dass sie auf einem angegriffenem System keinen Schaden anrichten.

```
alert ip $HONEYNET any -> $EXTERNAL_NET any
(msg:"SHELLCODE x86 stealth NOOP"; rev:6; sid:651;
content:"|EB 02 EB 02 EB 02|";
replace:"|24 00 99 DE 6C 3E|";)
```

Abbildung 2: Snort_inline-Regel zur Modifikation von Paketen mit x86-Shellcode

Die Modifizierung sorgt im Gegensatz zur Blockierung von Paketen dafür, dass der Angreifer versuchen kann, andere Systeme anzugreifen, aber dennoch scheitert. Dies ermöglicht, die Vorgehensweise eines Angreifers so gering wie möglich zu beeinflussen, jedoch ohne dass von ihm weiterer Schaden ausgeht. Mittels Snort_inline kann man das Risiko eines Angriffs vom *honeynet* auf andere Systeme zwar verringern, dennoch besteht aber natürlich weiterhin die Gefahr, dass mittels bisher unbekannter Schwachstellen oder veränderter *exploits* andere Systeme angegriffen werden. Dieses Restrisiko bleibt beim Einsatz eines *honeynet* naturgemäß bestehen.

Außer der Steuerung des Datenflusses hat die *honeywall* die Aufgabe, für die Speicherung aller gesammelter Informationen innerhalb des *honeynet* zu sorgen. Dazu wurde auf diesem System der Sebek-Server installiert, der die von den Sebek-Clients gesendeten Datenpakete sammelt und in eine Datenbank schreibt. Mittels Sebek können sämtliche Daten, die von den *honeypots* verarbeitet werden (insbesondere auch verschlüsselte SSH- oder HTTPS-Verbindungen), aufgezeichnet und später analysiert werden. Zusätzlich sorgen die Speicherung aller von Snort_inline gesammelten Datenpakete in einer Datenbank und die Log-Dateien von `iptables` für eine gewisse Redundanz der gesammelten Informationen.

Darüber hinaus wurden einige Programme wie etwa ACID [aci] oder Swatch [swa04] installiert, um die gesammelten Daten leichter zu verarbeiten und eine Alarmierung bei entdeckter Kompromittierung auszulösen. Außerdem wurde eine Härtung des Systems durchgeführt, um das Risiko eines Angriffs auf die *honeywall* zu minimieren.

## 4 Statusbericht

Das Aufsetzen des *honeynet* verlief weitgehend reibungslos. Lediglich ein automatisches Laden des Sebek-Client-Moduls beim Systemstart und anschließendes „Verstecken" dieses Moduls bereitet noch Probleme. Dies soll allerdings durch ein Umschreiben des Sebek-Clients gelöst werden. Insgesamt erforderte die Realisierung der in Abbildung 1 dargestellten Konfiguration weniger als eine Personenwoche Arbeitszeit. Beim Testen der Unsichtbarkeit des Sebek-Client-Moduls wurde desweiteren eine einfache Möglichkeit entdeckt, mit der ein *blackhat* selbst ohne privilegierte Rechte die Existenz von Sebek auf dem *honeypot* entdecken kann. Der Entwickler des Sebek-Clients wurde darüber informiert. Eine systematische Analyse des Fehlers findet im Moment statt.

Eine größere Hürde bildeten verständlicherweise die für Netzwerksicherheit verantwortlichen Mitarbeiter des Rechenzentrums der RWTH Aachen, die den ein- und ausgehenden Netzwerkverkehr zentral kontrollieren. Die dort anfänglich vorherrschende Skepsis ist natürlich verständlich, da ein *honeynet* auch Gefahren für die Netzinfrastruktur und für andere Computer innerhalb des Netzwerkes in sich birgt. Nach einer erfolgten Kompromittierung eines *honeypot* kann ein Angreifer versuchen, von dort aus weitere Systeme im internen Netz der RWTH anzugreifen. Generell gelten Systeme innerhalb des gleichen Subnetzes als einfacher angreifbar, da in der Regel keine Firewall den Netzwerkverkehr filtert. Snort_inline sorgt zwar dafür, dass bereits bekannte Angriffe vereitelt werden, allerdings liefert dies eben keinen garantierten Schutz.

Durch konstruktive Gespräche konnte allerdings eine für beide Seiten annehmbare Lösung gefunden werden. Dem *honeynet* wurde ein Subnetz mit 64 IP-Adressen zugegewiesen. Eine dedizierte Leitung vom Rechenzentrum hin zum *honeynet* ermöglicht den reibungslosen Betrieb und sorgt für eine Absicherung des internen Netzes der RWTH, da keine physische (Kabel-)Verbindung in dieses besteht.

Seit Mitte Februar 2004 ist der Linux-*honeypot* vom Test- in den Produktionsbetrieb übergegangen und sammelt seitdem Daten über den Netzwerkverkehr. Bisher fand noch keine

Kompromittierung des *honeynet* statt, allerdings gewährt eine Auswertung der bis zum 15. April 2004 aufgezeichneten Daten bereits interessante Rückschlüsse auf die Aktivitäten von *blackhats*:

- In den ersten beiden Monaten wurden etwa 61 MB an Netzwerkverkehr mitgeschnitten; dies entspricht etwa 425.000 Datenpaketen. Insgesamt wurden mehr als 9.500 verschiedene IP-Adressen ermittelt, die Datenpakete zu den *honeypots* verschickten; dabei wurden vermutlich auch per *IP spoofing* gefälschte Adressen beobachtet.

  Auch die passive Existenz eines Computersystems im Internet, d.h. eine Anbindung ohne ausgehenden Netzwerkverkehr, lässt es also durchaus zum Angriffsziel werden.

- Der Zugriff über das Netz fand schon etwa zehn Minuten nach dem Anschluss des *honeynet* an das Internet statt. Das System wurde systematisch auf Schwachstellen hin untersucht (*port scan*). Ein ungepatchter Internet Information Server (IIS) von Microsoft wäre nach dieser kurzen Zeitspanne schon kompromittiert gewesen.

- Es wurden vor allem *scans* nach `cmd.exe` durchgeführt. In diesen beiden ersten Monaten fanden knapp 10.000 *scans* dieses Typs statt. Dies deutet auf eine immer noch hohe Aktivität von *script kiddies* und Würmern hin, die nach den Hintertüren von *Code Red* oder Verwundbarkeiten des Internet Information Server suchen. *Code Red* selbst scheint in verschiedenen Versionen noch aktiv zu sein, ca. 1.500 mal wurde ein *scan* nach einer solchen Schwachstelle registriert.

- Die restlichen vom *honeynet* angebotenen Dienste wie FTP, SSH und PHP-Nuke wurden nur selten beachtet. Lediglich ein *scan* nach FTP-Servern, die jederman das Ablegen von Daten erlauben, konnte mehrfach beobachtet werden. Bei den beiden anderen Diensten gab es jeweils nur weniger als ein Dutzend Verbindungsversuche.

Desweiteren konnte gezeigt werden, dass es für einen versierten Angreifer möglich ist, Sebek zu entdecken, zu deaktivieren und zu umgehen [DHK04].

# 5 Ethische und Rechtliche Aspekte

Dieser Abschnitt bietet einen Überblick über die ethischen Aspekte und die rechtlichen Rahmenbedingungen, die beim Betrieb eines *honeynet* insbesondere in Deutschland beachtet werden müssen. Dies kann natürlich keine umfassende juristische Analyse ersetzen, da diese sehr stark vom konkreten Einzelfall abhängt, reicht aber für eine erste Einschätzung des juristischen Risikos.

Die Verantwortlichkeit des *honeynet*-Betreibers ist in zwei Bereichen besonders diskussionswürdig:

- Wie ist das *honeynet* in Bezug auf das gesamte Internet und wie sind insbesondere Angriffe von *honeypots* aus auf andere Systeme zu beurteilen? Hierbei sind ethische, straf- und zivilrechtliche Aspekte von Interesse.

- Wie ist es zu bewerten, dass die *blackhats* ohne ihr Wissen zum Teil eines Experimentes gemacht werden? Hierbei kommen insbesondere datenschutzrechtliche Aspekte zum Tragen.

### 5.1 Ethische Aspekte

Durch die Installation eines mit dem Internet verbundenen *honeynet* wurden dem Internet Systeme hinzugefügt, die wider besseren Wissen nicht dem Stand der Sicherheitstechnik entsprechen. Man kann argumentieren, dass dadurch grundsätzlich die Gesamtsicherheit des Internets verringert wurde. Für den *honeynet*-Betreiber würde dies bedeuten, dass er eine besondere Verantwortung gegenüber Personen übernimmt, deren Systeme durch *blackhats* von dem *honeynet* aus angegriffen werden. Betrachtet man jedoch das aktuelle Sicherheitsniveau von Systemen im Internet, scheint dieses dermaßen niedrig zu sein, dass die *honeypot*-Systeme möglicherweise sogar *über* dem durchschnittlichen Sicherheitsniveau liegen und somit die Gesamtsicherheit des Internet *verbessern*.

Bedenkt man weiterhin, dass (1) die Systeme des *honeynet* mit ihrer mangelnden Sicherheit die Population potentieller Opfer im Internet vergrößert und damit das Risiko des einzelnen verringert, Opfer einer Attacke zu werden, und (2) gleichzeitig auch durch die strenge Kontrolle des ausgehenden Verkehrs der Anteil an Rechnern verringert wird, die ohne weiteres als Plattform für weitere Angriffe genutzt werden können, kann durchaus davon ausgegangen werden, dass das *honeynet* allein durch seine Existenz das Risiko dritter, zum Opfer eines Angriffs aus dem Internet zu werden, leicht verringert. Weiterhin besteht die berechtigte Hoffnung, dass die mit dem *honeynet* gesammelten Ergebnisse mittelfristig zur Steigerung der Gesamtsicherheit des Internet beitragen. Ethisch scheint daher der Betrieb eines *honeynet* gegenüber der Gesellschaft vertretbar zu sein.

### 5.2 Strafrechtliche Aspekte

Rechtlich könnte eine straf- oder zivilrechtliche Haftung bestehen, wenn das *honeynet* Teil eines Angriffs gegen Dritte ist. Strafrechtlich könnte der Betrieb eines *honeynet* gemäß § 27 StGB Beihilfe zu Straftaten sein, die von einem *blackhat* über das *honeynet* verübt werden. Beihilfe kann jedoch nur durch eine *vorsätzliche* Hilfeleistung erfolgen. Dies ist nicht gegeben. Stattdessen wird ja versucht, mittels Snort_inline und `iptables` Beeinträchtigungen anderer Systeme von dem *honeynet* aus zu verhindern. Weiterhin ist der Betrieb eines Rechners mit dem Sicherheitsniveau der Honeypots völlig sozialadäquat, d.h. er lädt nicht etwa *blackhats* zum Missbrauch ein. Der Sicherheitsstandard der Systeme liegt sogar über dem vieler anderer am Internet angeschlossener Systeme. Strafrechtlich ist der Betrieb des *honeynet* daher unbedenklich.

### 5.3 Zivilrechtliche Aspekte

Eine zivilrechtliche Haftung für durch den Angreifer vom *honeynet* aus gegenüber anderen verursachten Schäden käme primär aus § 823 I BGB in Betracht. Dafür müsste aber ein von dem *blackhat* verursachter Schaden den Betreibern des *honeynet* zurechenbar sein. Da die Betreiber des *honeynet* jedoch nicht aktiv andere Rechner schädigen, kommt nur eine Haftung aus dem Unterlassen der Absicherung des *honeynet* in Betracht. Dazu müsste der Betreiber insbesondere eine Garantenstellung innehaben, die ihn verpflichtet, Gefahren aus dem *honeynet* einzuschränken. Diese könnte sich aus einer Verkehrssicherungspflicht ergeben. Die allgemeine Verkehrssicherungspflicht besagt, dass jedermann, der in seinem Verantwortungsbereich Gefahren schafft oder andauern lässt, die notwendigen Vorkehrungen treffen muss, die im Rahmen des wirtschaftlich zumutbaren geeignet sind, Gefahren von Dritten abzuwenden.

Welche Maßnahmen notwendig und zumutbar sind, ist eine Wertungsfrage. Bisher scheint es die Gesellschaft inklusive der Rechtsprechung abzulehnen, Betreiber von mit dem Internet verbundenen Systemen für Schäden zur Verantwortung zu ziehen, die durch die Unsicherheit dieser Systeme entstanden sind. Daher ist davon auszugehen, dass es als nicht notwendig oder als nicht zumutbar angesehen wird, mit dem Internet verbundene Systeme abzusichern oder zu überwachen. Das *honeynet* wird jedoch umfangreich überwacht. Insbesondere wird besonderer Aufwand betrieben, um Schaden von Dritten abzuwenden. Damit hat der Betreiber des *honeynet*, selbst wenn man eine Verkehrssicherungspflicht für mit dem Internet verbundene Systeme annähme, diese erfüllt.

### 5.4 Datenschutzrechtliche Aspekte

Bei der Frage, wie der Betrieb des *honeynet* gegenüber den eindringenden *blackhats* zu bewerten ist, ist vor allem fraglich, in wie weit ein angreifender *blackhat* nicht unwissend zum Teil eines Experiments gemacht wird. Der *blackhat*–Community ist durchaus bewusst, dass zum einen *honeynets* existieren und dass zum anderen Systemadministratoren soweit sich entsprechende Gelegenheiten bieten, alle Schritte von *blackhats* aufzeichnen und umfangreiche forensische Untersuchungen durchführen. Deshalb kann davon ausgegangen werden, dass *blackhats* billigend in Kauf nehmen, dass ihr Handeln aufgezeichnet und untersucht wird. Auch rechtlich bestehen aus dieser Hinsicht keine Bedenken. Insbesondere ist nicht davon auszugehen, dass das *honeynet* mit personenbezogenen Daten in Kontakt kommt. Somit ist der Betrieb aus datenschutzrechtlicher Sicht unbedenklich.

## 6  Zukünftige Arbeiten

Nach einer Testphase mit dem in Abschnitt 3 beschriebenen *honeynet* soll in einer folgenden Phase eine Erweiterung um ein virtuelles *honeynet* auf Basis von User-mode Linux [use04] oder VMware [vmw04] erfolgen. Bei einem virtuellen *honeynet* werden virtuelle

Maschinen benutzt, um *honeypots* aufzusetzen. So erreicht man eine einfachere Administration und Wartbarkeit, da man auf einem Computer mehrere *honeypots* verwalten kann.

User-mode Linux [use04] ist eine Portierung von Linux auf Linux, die es ermöglicht, mehrere Instanzen des Linux-Kernels auf einem Linux-Rechner zu starten. Damit kann man ohne großen Aufwand mehrere *honeypots* simulieren. Allerdings ist man mit dieser Lösung auf Linux als simuliertes Betriebssystem beschränkt. Dies kann man durch Verwendung von VMware [vmw04] vermeiden, einem kommerziellen System zur Emulation von Computern. Mit VMware kann auf den virtuellen Maschinen unter anderem Microsoft Windows (Windows 3.1x bis Windows Server 2003), Linux (Kernel-Versionen 2.2, 2.4 und 2.6.) und Novell NetWare betrieben werden. Allerdings sind hierbei die Anforderungen an die Hardware wesentlich höher und die Lizenzkosten relativ hoch. Aufgrund der erhöhten Aktivität von *blackhats* in Bezug auf den Internet Information Server von Microsoft scheint die Einrichtung eines *honeypot* mit Microsoft Windows eine sinnvolle Erweiterung des *honeynet* zu sein. Daher wird dies als nächstes angestrebt.

Abbildung 3 zeigt die geplante Erweiterung um virtuelle Maschinen. Die genaue Zusammensetzung des virtuellen *honeynet* ist noch in Planung, und deshalb ist noch keine Aussage zur genauen Konfiguration oder den verwendeten Plattformen möglich.

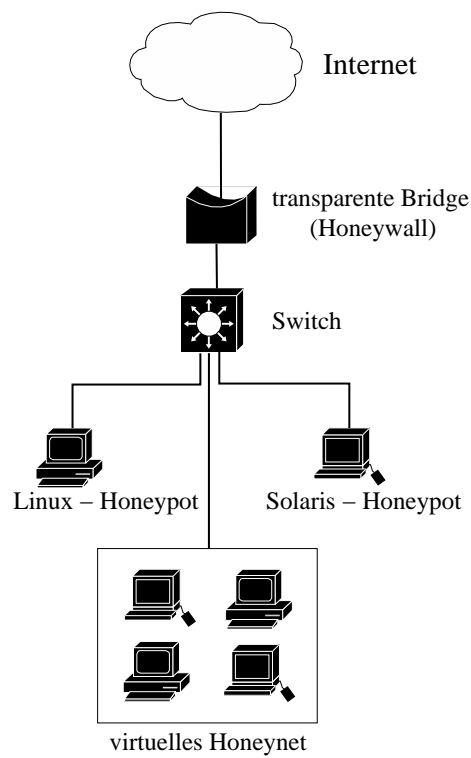

Abbildung 3: Geplante Konfiguration des *honeynet* an der RWTH Aachen.

Außer der Erweiterung des bestehenden *honeynet* um virtuelle *honeypots* sollen in nächster Zeit Programme entwickelt werden, mit denen die Analyse der gesammelten Daten vereinfacht werden kann. Es existiert zwar ein Web-Interface, mit dessen Hilfe die Daten von Sebek ausgewertet werden können, aber eine Weiterentwicklung dieses System sowie die Erstellung weiterer Programme zur Datenanalyse wird angestrebt.

Zudem ist auch eine Weiterentwicklung der schon bestehenden Programme zur Datensteuerung und -sammlung (insbesondere Snort_inline und Sebek) angedacht. Die Veröffentlichung des Quellcodes der verwendeten Programme führt nämlich dazu, dass *blackhats* Methoden entwickeln können, um *honeynets* zu entdecken. Dies ist in der Vergangenheit mehrmals dokumentiert worden [Co03, Co04]. Gleichzeitig führt die Weiterentwicklung von *rootkits* dazu, dass in Sebek immer neue Ideen einfließen, um das Programm noch transparenter in das System zu integrieren. Es findet also ein ständiger Wettlauf um die neuesten Methoden zur Entdeckung und Verschleierung von *honeynets* statt, der eine Weiterentwicklung auf diesem Gebiet unerlässlich macht.

Die gegenwärtig bei *honeynets* verwendete Methodik schränkt das untersuchbare Verhalten der *blackhats* sehr ein. Es werden praktisch nur Angriffe auf mäßig gesicherte Unix-Serversysteme ohne Anwendungen mit komplexen Benutzerschnittstellen und ohne aktive Nutzung untersucht. Weiterhin können praktisch nur „zufällige" Angreifer detektiert werden: Es handelt sich also in der Regel entweder um automatisierte Angriffe von Würmern und dergleichen oder um Angriffe von *blackhats*, die ihre Ziele mehr oder weniger nach dem Zufallsprinzip aussuchen. Dass ein *blackhat*, der seine Ziele planmäßig auswählt, Systeme im *honeynet* angreifen wird, scheint unwahrscheinlich.

Die Frage, wie die Bereiche der abgedeckten *blackhat*-Aktivitäten ausgedehnt werden können, muss weiter diskutiert werden. Ein erster Ansatz besteht darin, auf den *honeypots* Software wie PHP-Nuke [php04] zu installieren, die es ermöglicht, Angriffe auf Webapplikationen zu beobachten. Problematisch ist hierbei allerdings die Frage, ob und in welcher Hinsicht die Simulation von Benutzeraktivität sinnvoll ist. Wie mit Hilfe von *honeynet*-Technologie andere Aktivitäten beobachtet werden können, insbesondere Ressourcendiebstahl in Form von *spam relaying* oder Proxy-Mißbrauch und Angriffe auf Desktop Systeme in Form von Viren, Würmern, *spyware*, Remote Access Trojanern und Dialern, ist noch ungeklärt.

**Danksagungen**

# Literatur